\documentstyle[12pt]{article}
\title{Galilean-Invariant $(2+1)$-Dimensional Models with a Chern-Simons-Like
Term and $D=2$ Noncommutative Geometry}
\author{Jerzy Lukierski\thanks{Department of Mathematical Sciences, Science
 Laboratoires, University od Durham, South Road, Durham DH1 3LE, England}
\thanks{On leave of absence from the Institute of Theoretical Physics,
University of Wroc\l{}aw, pl. Maxa Borna 9, 50-204 Wroc\l{}aw, Poland} 
\and Peter C. Stichel\thanks{Faculty of Physics, University of Bielefield,
Universit\"atsstr.25, 33615 Bielefeld,
Germany} \and  Wojtek J. Zakrzewski\footnotemark[1] }  \date{}
\newcounter{popnr}
\def\theequation{\thesection.\arabic{equation}}
\renewcommand{\theequation}{\arabic{section}.\arabic{equation}}
\newcommand{\alpheqn}{\setcounter{popnr}{\value{equation}}
                      \addtocounter{popnr}{1}
                      \setcounter{equation}{0}
   \renewcommand{\theequation}{\arabic{section}.\arabic{popnr}\alph{equation}}}
\newcommand{\reseteqn}{\setcounter{equation}{\value{popnr}}
     \renewcommand{\theequation}
     {\arabic{section}.\arabic{equation}}}
\def\poin{Poincar\'e}
\def\ddd{\stackrel{\dots}{x}}
\def\annexe#1#2{\def\thesection{\Alph{section}}\section*{#2}
                \setcounter{section}{#1}  }
\def\dddd{\stackrel{\dots .}{x}}
\newcommand{\be}{\begin{equation}}

\def\lbl{\label}
\def\implies{\Longrightarrow}
\def\F{{\cal F}}
\def\tQ{{\widetilde Q}}
\def\bl{\alpheqn}
\def\el{\reseteqn}
\def\bel{\begin{equation}\label}
\def\ee{\end{equation}}
\def\beq{\begin{eqnarray}}
\def\eeq{\end{eqnarray}}
\def\ba{\begin{array}}
\def\ea{\end{array}}
\def\1{{\bf1}}
\def\r#1{(\ref{#1})}
\def\L{{\cal L}}
\def\Lam{\Lambda}
\def\lam{\lambda}
\def\a{\alpha}
\def\e{\epsilon}
\def\om{\omega}
\def\Om{\Omega}
\def\i{\item}
\def\d{\delta}
\def\ben{\begin{enumerate}}
\def\een{\end{enumerate}}
\def\kx{\dot x}
\def\kkx{\ddot x}
\def\kkkx{\mathop{x}\limits^{\ldots}{\!}}

\def\pa#1#2{\frac{\partial#1}{\partial #2}}
\def\dt#1{\frac{d #1}{d t}}
\def\p{\partial}
\def\sec{\setcounter{equation}{0}}
\def\tp{\widetilde p}
\def\tP{\widetilde P}
\def\ty{\widetilde y}
\def\k{\dot}
\def\kp{\dot p}
\def\ktp{\mathop{\widetilde p}\limits^{.}{\!}}
\def\ktP{\mathop{\widetilde P}\limits^{.}{\!}}
\def\({\left(}
\def\){\right)}
\def\[{\left[}
\def\]{\right]}
\def\<{\left<}
\def\>{\right>}

\def\i{\item}
\def\dsp{\displaystyle}

\begin{document}
\maketitle
\vspace*{-6mm}
\begin{abstract}
We consider a new $D=2$ nonrelativistic  classical mechanics model
providing via the Noether theorem the $(2+1)$-Galilean symmetry algebra
with two central charges: mass $m$ and the coupling constant $k$
of a Chern-Simons-like term. In this way we provide the dynamical
interpretation of the second central charge of the $(2+1)$-dimensional Galilean
algebra. We discuss also the interpretation of $k$ as describing the 
noncommutativity of $D=2$ space coordinates. The model is quantized 
in 
two ways: using the Ostrogradski-Dirac formalism for higher order Lagrangians 
with constraints and the Faddeev-Jackiw method which describes
 constrained systems and produces
 nonstandard symplectic structures. We show that our model describes
 the superposition of a free motion
in noncommutative $D=2$ space as well as the ``internal" oscillator modes.
We add a suitably chosen class of velocity-dependent two-particle 
interactions, which  is descrobed by local potentials
in  $D=2$ noncommutative space. We treat, in detail, the particular
 case of a harmonic
 oscillator and describe its quantization. 
It appears that the indefinite 
metric  due to the third order time derivative term in the field equations, even 
in the presence of interactions, can be eliminated by the imposition
of a subsidiary condition. 
\end{abstract}

\section{Introduction}

In a $N$-dimensional nonrelativistic classical mechanics the Galilean symmetry
transformations ($i,j=1,\ldots,N$)
\bel{1.1}
\ba{rcl}
x'_i&=& R_i{}^j x_j+v_it + a_i\,,\\
t'&=& t + \tau
\ea
\ee
generated by the Galilei algebra $G_N$ leave the equation of motion invariant,
but quite often the Lagrangian is changed under the transformations \r{1.1}
by a total time derivative (see e.g\ [1,2]). Such a quasi-invariance of the 
Lagrangian leads to the appearance of a central extension $G_N\to \hat G_N$ 
of the Galilean algebra. Let us recall that $G_N$ is described by $\frac12N(N-1)$
rotation generators $J_{ij}=-J_{ji}$ (generate $O(N)$ rotations $R_i{}^j$), $N$
momenta
$P_i$ (generate space translations $a_i$), $N$ Galilean boosts $K_i$ (generate
velocities $v_i$) and the energy operator $H$ (generates time shifts $\tau$).
The best known central extension, occuring for any $N\geq1$, describes the mass
generator $M$ which modifies the commutativity of boosts and momenta as follows
(see [3])
\bel{1.2}
[K_i,P_j]=i\delta_{ij} M\,.
\ee
The relation \r{1.2} implies that for $M \ne 0$, by defining
\bel{1.3}
X_i=\frac{K_i}{M},
\ee
one can embed the Heisenberg algebra\footnotemark[1]
\bel{1.4}
[X_i,P_j]=i\delta_{ij}
\ee
into the enveloping algebra of $\hat G_N$. This 
property of nonrelativistic systems has important
consequences; in particular the no-interaction theorems valid for relativistic
\poin-invariant two-particle systems (see [2,4,5]) are not true in the
nonrelativistic case.

The simplest way of demonstrating the physical interpretation of the central
charge $M$ involves considering a free nonrelativistic particle, with the
Lagrangian $L_0=\frac12 m\kx_i^2$. Introducing the momenta $p_i=\pa{L_0}{\kx_i}
=m \kx_i$ we find from the Noether theorem applied to the transformations
\r{1.1} ($\vec p=(p_1\ldots p_N)$) that 
\bel{1.5}
\ba{c}
J_{ij}=x_ip_j-p_ix_j\,,\qquad P_i=p_i\,,\\
K_i = mx_i\,,\qquad H=\frac{\vec p^2}{2m}\,,\qquad M=m.
\ea
\ee
If we introduce the canonial commutation relations \r{1.4} for $X_i=x_i$ and
$P_j=p_i$ we can show that \r{1.5} povides the one-particle realization of the
Galilei algebra $G_N$, centrally extended by the mass generator $M=m\cdot 
\1$.Using the field equations $\kp_i=0$ we obtain further that the generators
$J_{ij}$, $P_i$ and $H$ are constant in time, and $K_i$ depend on time in
accordance with the Galilei algebra relation
\bel{1.6}
\dot K_i = i[H,K_i] =  P_i\,.
\ee
Let us add that it is the cohomological consideration (see e.g. [6]) which
shows that in three space time dimensions ($N=3$) the mass generator $M$ is the
only central charge which can be added to the ten generators of the classical
Galilei algebra $G_3$. This conclusion is not valid for $N=1$ and $N=2$;
for $N=1$ (one space, one time) we can introduce two central charges and for
$N=2$ (two space, one time) we have the possibility of three central
charges (the mass
$M$ and two additional central charges $K$, $E$  --- see [7]). In  the latter
case we have the following  extended Galilei algebra $\hat G_2$
($J_{12}=J$; $i,j=1,2$):
\bel{1.7}
\ba{rclrcl}
[J,K_i]&=&i\e_{ij}K_j\,,&{}\quad[J,P_i]&=&i\e_{ij}P_j,\phantom{aassssssssssaaaa}\,\\
{}[J,H]&=&iE\,,&{}\quad[K_i,P_j]&=&i\delta_{ij}M\,,\phantom{aaaaassssaaaaaaa}\\
{}[K_i,K_j]&=&i\e_{ij}K\,,&{}\quad[K_i,H]&=&iP_i,\quad[P_i,H]=[P_1,P_2]\,=\,0\,.
\ea
\ee
\setcounter{equation}{7}
Taking into consideration the mass dimensions of the generators $[K_i]=0$,
$[J]=0$,
$[P_i]=[H]=1$ we see that the central generators have dimensions $[M]=[E]=1$
and $[K]=0$. In what follows we shall restrict ourselves to the case 
$E=0$ because, as shown by Levy-Leblond [7], if $E\neq 0$ the algebra \r{1.7} 
can not be integrated to the extended $N=2$ Galilei group $E\neq0$. Indeed,
assuming that $E=e\cdot\1$, the finite $O(2)$ rotations are generated by $J$ 
 as follows:
\bel{1.8}
e^{\theta J} H e^{-\theta J} = H + e \theta 1\,.
\ee
However, as $\theta=2\pi$ and $\theta=0$ should give identical results, 
one can integrate the Lie algebra $\hat G_2$ to the corresponding Lie group
only if $e=0$.

The representations of the Lie algebra \r{1.7} with three central charges 
$M$, $K$, $E$ as well as the projective representations of the corresponding
$N=2$ Galilei group were studied by several authors (see e.g. [7-12]).
Indeed, in accordance with the general scheme (see e.g.\ [3,6]) the 
appearance of central charges in a Lie algebra leads, on the group level,
to the appearance of projective representations of the corresponding Lie group.

The main result of this paper involves finding a Lagrangian model
which provides, via the Noether theorem, the $N=2$ Galilean algebra $\hat G_2$ 
with two central charges $M=m\cdot \1$ and $K=2k\cdot \1$. The interest of 
having such a model is threefold:
\ben
\i[i)] One obtains a clear physical interpretation of the second central 
charge of $\hat G_2$
\i[ii)] If we keep  relation \r{1.3} the model possesses 
noncommutative space coordinates, i.e.\footnotemark[2]
\bel{1.9}
[X_i,X_j]=i\frac{k}{m^2}\epsilon_{ij}
\ee
\i[iii)] It provides a good example of the Faddeev-Jackiw
quantization method.
\een
As we shall show, our model can be described either in terms of phase space 
variables with commuting space coordinates, or in terms of new phase space 
variables with noncommutative space coordinates given by  relations 
\r{1.9}. After considering
free motion in the noncommutative space  we shall 
 introduce  interactions in the classical $D=2$ space 
generating  a potential term which depends on noncommuting 
$D=2$ space coordinates.
Recently, there have been
several proposals for deformations of space-time variables leading to their
noncommutativity (see e.g.\ [15-18]) and also to deformed classical
and quantum mechanics (see e.g. \ [19,20]). In our case we exploit the 
explicit relation between commuting and noncommuting position variables 
 and we expect that our model can 
contribute also to a better understanding of dynamical models on other noncommutative
spaces.

 The plan of our 
presentation is as follows. In Sect. 2 we present our model in a Lagrangian 
as well as Hamiltonian formulation, write down the corresponding constraints,
Dirac brackets and introduce the corresponding Ostrogradski-Dirac 
symplectic formalism. 
In the Hamiltonian formalism, following the scheme for higher order 
Lagrangians [21-24] one introduces
 besides the positions $x_i$ also two pairs of 
momenta $p_i$, $\widetilde p_i$ as phase space variables. It should be 
mentioned that an equivalent formalism is also possible, with the canonical 
variables ($x_i$, $\dot x_i$, $p_i$) and 
in Sect.\ 2 we shall show that the quantization with this choice of  
variables can be easily achieved using the 
geometrically motivated Faddeev-Jackiw method [13,14].
In Sect.\ 3, using the Ostrogradski phase space 
formalism and the Faddeev-Jackiw method
we discuss the Galilean symmetries: Noether charges and 
conservation laws. In Sect 4 we present the symplectic formalism for the 
choice of phase space with noncommuting space coordinates satisfying  
relations \r{1.9}, and use this framework to consider the dynamics of the 
model. We arrive at the conclusion that the Hamiltonian of section 2 can be diagonalised and
 that it
describes a free motion in the noncommutative phase space supplemented 
by the oscillator modes with negative sign of their energies.
In sect. 5 we introduce the two particle $D=2$ Galilean invariant dynamics 
and consider the class of velocity-dependent interactions, which imply
the appearance of a potential term in the noncommutative $D=2$ space.
In particular, we study in detail a model with noncommutative
harmonic forces, describing a harmonic oscillator in the $D=2$ 
noncommutative space which was first
introduced, in the Hamiltonian framework, in [10]. We find that although the 
parameter$k$ (see (1.9)) modifies the standard spectrum of the oscillator 
all its eigenvalues remain positive. In sect. 6 we discuss the problem of 
indefinite metric. We find that the modes carrying indefinite metric can 
always be removed by the imposition of a Gupta-Bleuler type of
a subsidiary 
condition. The paper contains also an Appendix in which
we show that our Lagrangian is the most general $D=2$  Galilei-invariant
Lagrangian linear in the second time derivative of the position variable.

\sec
\section{A model with a Chern-Simons-like term}

As is well known, in two dimensions, due to the existence of the Levi-Civita 
antisymmetric metric $\e_{ij}$, one can introduce a free particle action
with a Chern-Simons like term ($\lambda$ has  dimension of mass/time):
\bel{2.1}
{\L} = \frac{m\dot x^2}{2} + \lambda \e_{ij} x_i \dot x_j\,.
\ee
The second term can be interpreted as a coupling $\lam A_j \dot x_j$ of a 
particular electromagnetic potential $A_j=\e_{ij} x_i$ corresponding 
to constant magnetic field strength $F_{ij}=\p_i A_j - \p_j A_i
=\lam \e_{ij}$. The Lagrangian \r{2.1} is neither invariant nor 
invariant up to a total derivative
under the Galilean boost transformations; the symmetry 
algebra is described by the Hamiltonian $H$ and the $D=2$ Euclidean inhomogeneous 
algebra ($J$, $P_1$,$P_2$) centrally extended by the central charge 
$\Lam=\lam\cdot \1$:\bel{2.2}
[J,P_i]= i\e_{ij} P_j\,,\qquad [P_i,P_j]=2i\e_{ij}\Lam\,.
\ee

In order to obtain a two-dimensional model which is quasi-invariant under 
$D=2$ Galilei symmetry we modify the second term in \r{2.1} and consider 
($k$ has the physical dimension of mass $\times$ time)
\bel{2.3}
L= \frac{m\kx^2}{2} - k \e_{ij} \kx_i \kkx_j\,.
\ee

It is interesting to observe that following the methods of [25] one can show
that the Lagrangian (2.3) is the most general one which is quasi-invariant
under the $D=2$ Galilei transformations and which contains at most a linear dependence on the second derivative terms $\ddot x_i$ (see the Appendix).

\subsection*{A) Quantization using the Ostrogradski-Dirac method}

The Hamiltonian description of the Lagrangian \r{2.3} follows 
from the Ostrogradski formalism for higher order Lagrangians, supplemented  
by the Dirac bracket technique. Due to the presence of a second order 
derivative in the Lagrangian we have to introduce two momenta:
\bl
\bel{2.4a}
p_i=\pa{L}{\kx_i} -\dt{}{\partial L\over \partial \ddot x_i} \,\qquad \tp_i =\pa{L}{\kkx_i}\,.
\ee
Hence in our case 
\bel{2.4b}
p_i = m\kx_i - 2 k \e_{ij} \kkx_j\,. \qquad \tp_i 
= k\e_{ij} \kx_j\,.
\ee
\el

The Lagrange equation of motion 
\bel{2.5}
\pa{L}{x_i}-\dt{}\pa{L}{\kx_i}+\frac{d^2}{dt^2}\pa{L}{\kkx_i}=0
\ee
takes in our case the form 
\bel{2.6}
\kp_i = m\kkx_i - 2k\e_{ij}\kkkx_j=0\,.
\ee
The Hamiltonian is given by
\bel{2.7}
H=\kx_i p_{i} + \kkx\tp_i -L =
\frac{m\kx_i^2}{2} - 2k\e_{ij}\kx_i\kkx_j\,.
\ee
Because
\bl
\bel{2.8a}
\frac{m\kx_i^2}2 =\frac{m}{2k^2}(\tp_j)^2
\ee
\bel{2.8b}
\e_{ij}\kx_i\ddot x_j=-\frac1{2k^2}\tp_k \e_{kl}p_l +\frac{m}{2k^3}(\tp_j)^2\,,
\ee
\el
we obtain
\bel{2.9}
H=-\frac{m}{2k^2}(\tp_j)^2 +\frac1k \tp_k\e_{kl}p_l\,.
\ee

The Hamiltonian formalism for the Lagrangian \r{2.3} can be written in the 
eight-dimensional phase space ($x_i$, $\kx_i$, $p_{i}$, $\tilde p_{i}$) with 
two constraints
\bel{2.10}
\Phi_i = \kx_i+\frac1k \e_{ij}\tp_j=0\,.
\ee

These constraints lead to the replacement
of the canonical Poisson brackets 
\begin{eqnarray}
\nonumber \{x_i,p_j\}&=&\delta_{ij}\,\qquad \{\kx-i,\tp_j\}\,=\,\delta_{ij}\,,\\
\label{2.11} \{x_i,\kx_j\}&=&\{p_i,\tp_j\}\,=\,0\,,\\
\nonumber \{x_i,\tp_j\}&=&\{\kx_i,p_j\}\,=\,0\,,
\end{eqnarray}
 by the Dirac brackets
\bel{2.12}
\{X,Y\}_D =\{X,Y\} - \{X,\Phi_i\} \frac k 2 \e_{ij} \{\Phi_j,Y\}\,,
\ee
where $\frac k 2 \e_{ij} =C_{ij}^{-1}$ and $C_{ij}=\{\Phi_i,\Phi_j\}$. 
In particular, the fundamental Poisson bracket relations are replaced by the symplectic structure depending on the choice of six independent
canonical variables. We have two possibilities:

\ben
\i[i)] The phase space with two sets of momenta $y_A=(x_i,p_i,\tp_i)$.\\
Then from \r{2.12} we have
\bel{2.13}
\{y_A,y_B\}_D=\omega _{AB}\,,
\ee
where
\bel{2.14}
\omega=\pmatrix {0&\1_2&0\cr
-\1_2&0&0\cr
0&0&\frac{k}2\e}\,.
\ee
The Hamiltonian equations of motion
\bel{2.15}
y_A=\{y_A,H\}_D\,
\ee
where $H$ is given by \r{2.9}, take the form:
\bl
\beq
\label{2.16a} \kx_i&=&\{x_i,H\}_D\,=\,-\frac 1 k \e_{ij} \tp_j\,,\\
\label{2.16b} \kp_i &=& \{p_i,H\}_D \,=\,0\,,\\
\label{2.16c} \ktp_i&=& \{\tp_i,H\}_D \,=\, -\frac{m}{2k}\e_{ij}\tp_j -\frac12 p_i\,.
\eeq
\el
Substituting the constraint equation \r{2.16a} into \r{2.16c}
differentiating and using \r{2.16b} reproduces the equations \r{2.6}. 

To obtain the quantized form of the canonical commutation relations \r{2.13}
as well as the Heisenberg equations of motion we perform the replacement
\bel{2.17}
\{y,y'\}_D \to \frac1{i\hbar}[\hat y,\hat y']\,,
\ee
where $\hat y$, $\hat y'$ denote the quantized variables.

\i[ii)] The choice of independent variables $\ty_A=(x_i,\kx_i,p_i)$.\\
The symplectic structure is given by 
\bel{2.18}
\{\ty_A,\ty_B\}_{\tilde D}=\widetilde\om_{AB}\,,
\ee
where
\bel{2.19}
\widetilde\om=\pmatrix{0&0&\1_2\cr
0 & -\frac1{2k}\e & 0\cr
-\1_2 &0&0}\,.
\ee
The Hamiltonian $ H$ reads  using \r{2.10}
\bel{2.20}
H=\,\dot x_j\,p_j\,-\,{m\dot x_j\sp2\over 2}\,.
\ee
The Hamiltonian equations are
\bl
\beq
\label{2.21a} \ddot x_i&=&\{\dot x_i,H\}_{\tilde D}\,=\,-{m\over 
2k}\epsilon_{ij}\dot x_j\,+\,{1\over 2k}\epsilon_{ij}p_j\,,\\
\label{2.21b} \dot p_i&=&\{p_i,H\}_{\tilde D}\,=\,0\,,
\eeq\
\el
and the relation $\dot x_i=\{x_i,H\}_{\tilde D}$ has become an identity.
One can easily see that the equations \r{2.16a}-\r{2.16c} and 
\r{2.21a}-\r{2.21b} supplemented by the constraints \r{2.10} are 
equivalent.

\een
\subsection*{b) Quantization using the Faddeev-Jackiw method}

It appears that Lagrangians with higher order derivatives can be also 
treated by the Faddeev-Jackiw method [13,14];  particularly well-suited are 
the Lagrangians which are linear in the highest order derivatives. If we assume the 
Lagrangian of the form
\bel{2.22}
L(x_i,\dot x_i,\ldots,\mathop{x_i}\limits^{(n)}) = L^{(0)}(x_i,\,\ldots,
\mathop{x_i}\limits^{(n-1)})
+{L}_i^{(1)}(x_i,\,\ldots,\mathop{x_i}\limits^{(n-1)})
\mathop{x}_i\limits^{(n)},
\ee
then by introducing $n-1$ momenta ($p_i$,\,\ldots, $p_{i;k}$;\,$k=1$,\ldots 
$n-1$) as the Lagrange multipliers we can rewrite \r{2.22} as  
($i,j=1,\ldots d$)
\bel{2.23}
\ba{rcl}
L&=&\dsp L^{(0)} (x_j,y_{j;1},...,y_{j;n-1}) + 
     L_i^{(0)} (x_j,y_{j;1},...,y_{j;n-1}) \cdot \dot y_{i;n-1} \\[3mm]
 &&\dsp  + p_i (x_i-y_{i;1})+ \sum_{k=1}^{n-2}p_{i;k}
    (\dot y_{i;k}-y_{i;k+1})  \\[3mm]
 &=&\dsp p_i\dot x_i + \sum_{k=1}^{n-2} p_{i;k} \dot y_{i;k} + 
L^{(1)}_i(x_j,y_{j;1},... y_{j;n-1}) \cdot \dot y_{i;n-1} -H\,,
\ea
\ee
where
\bel{2.24}
H=  p_i y_{i;1} + \sum_{k=1}^{n-2}p_{i;k}y_{i;k+1} - 
      L^{(0)}(x_j,y_{j;1},... y_{j;n-1})\,.
\ee

The relation \r{2.23} is written in the form presented in [13, 14]; in 
particular, the canonical one-form determining the equal-time commutator 
algebra is given by
\bel{2.25}
A_K(Y) dY_K = p_i\,dx_i+\sum_{k=1}^{n-2} p_{i;k}dy_{i;k} +
L_i^{(1)} (x_j,y_{j;1},\ldots,y_{j;n-1})\,d y_{i;n-1}\,,
\ee
where $Y_K=(x_i,p_i,p_{i;1},y_{i;1},\ldots,p_{i;n-2},y_{i;n-2},y_{i;n-1})$.

If we introduce the antisymmetric tensor 
\bel{2.26}
f_{KL} = \frac{\p A_K}{\p Y_L} - \frac {\p A_L}{\p Y_K}\,.
\ee
and assume that the matrix \r{2.26} is invertible
we find that the basic Poisson brackets are given by
\bel{2.27}
\omega_{KL}=\{Y_K,Y_L\}=f_{KL}^{-1}\,.
\ee

Our Lagrangian \r{2.3} fits very well into such a scheme, which is obtained 
from\r{2.23} by putting $d=2$, $n=3$, $y_{i;1}\equiv y_i$, 
$Y_A=(x_1,x_2,p_1,p_2,y_1,y_2)$ and ($\vec y=(y_1,y_2)$)
\bel{2.28}
L^{(0)} = m\frac{\vec y^2}{2}\,,\qquad L^{(1)}_i= - k\epsilon_{ij}y_j\,.
\ee
We have 
\bel{2.29}
f_{KL} =\pmatrix{0 &  \bf 1_2 &0 \cr
                 -\bf 1_2 & 0 & 0 \cr
                 0 & 0 & -2k\, \epsilon_{ij} }\,,
\ee
{\em i.e.} from \r{2.27} we obtain the symplectic structure \r{2.19}. It 
should be noted that the Faddeev-Jackiw method, presented in this 
subsection, provides the quantization of the model \r{2.3} in an easier way 
than the conventional Ostrogradski-Dirac approach.

\sec
\section{Noether charges and the generalized $D=2$ Galilean algebra}
\subsection*{a) Ostrogradski-Dirac method}
\par Let us consider a Lagrangian $L(x_i,\kx_i,\kkx_i)$ which depends on the first
and second time derivatives. The variation of the action $S=\int dt L$ under
the change $ x_i\to x_i +\delta x_i$ takes the form (see e.g\ [21,22])
\bel{3.1}
\ba{rcl}
\delta S= \int dt\,\delta L &=& \int dt\,(\pa{L}{x_i} \d x_i + \pa L 
{\kx_i}\d\kx_i + \pa{L}{\kkx_i}\d\kkx_i)=\\[3mm]
&=& \int dt\,\dt{}(p_i \d x_i+ \tp_i \d\kx_i)
\ea
\ee
If $\d x_i = F_i^r (x_i,t) \d \a_r$ describes the symmetry, i.e.\
$\d L_{\a_r} =0$, we obtain the following formulae for the generator
\bel{3.2}
Q^r(t)=p_i(t) F_i{}^r(x_i(t),t)+\tp_i(t)\dt{}F_i{}^r(x_i(t),t)
\ee
which is conserved 
\bel{3.3}
\dt{}Q^r(t)=0 \implies Q^r(t_1)=Q^r(t_2)\,.
\ee
If the Lagrangian $L$ is quasi-invariant under the symmetry transformation, 
i.e\
\bel{3.4}
\d L_{\a_r} =\dt{}(\F^r \d\a_r)\,,
\ee
the generators \r{3.2} are not conserved. However, as the nonconservation law 
takes the form 
\bel{3.5}
\dt{}Q^r(t) = \dt{} \F^r\,,
\ee
we can introduce modified generators $\tQ^r=Q^r-\F^r$ which are conserved. 
The  generators $\tQ^r$ correspond to the modified symmetry 
transformations with central charges.

Let us list below the generators of the $D=2$ Galilei symmetry for the Lagrangian \r{2.3}.
We have 
\ben
\i[i)] translations ($\d x_i = \d \a_i$, $\d\kx_i=0$; $F_i{}^r = \d_i{}\sp{r}$,
$Q\sp{i}=P_i$)
\bel{3.6}
P_i =p_i
\ee
\i[ii)] rotations $O(2)$ ($\d x_i=-\e_{ij}x_j\d\a$, 
$\d\kx_i=-\e_{ij}\kx_j\d\a$; $F_i=-\e_{ij}x_j$, $Q=J$)
\bel{3.7}
J=x_i\e_{ij} p_j -\tp_i \e_{ij} \kx_j\,.
\ee
Using the constraint equations \r{2.10} we find that
\bel{3.8}
J=\e_{ij}x_i p_j -\frac1k (\tp_j)^2\,.
\ee
\i[iii)] Galilean boosts $\d x_i=v_i t$, $\d\kx_i=v_i$; $F_i{}^r=\d_i{}^r\cdot 
t$. (we denote the nonconserved Noether boost charges by $B_i$)

From \r{3.2} we obtain 
\bel{3.9}
B_i=p_i\cdot t +\tp_i\,.
\ee
\een

\noindent The Lagrangian \r{2.3} is quasi-invariant under Galilean boost 
transformations; 
in fact 
\bel{3.10}
\d L_{v_i}=\dt{}(mx_i-k\e_{ij}\kx_j)v_j
\ee
and the relation \r{3.5} takes the form 
\bel{3.11}
\dt{}B_i=
\dt{}(p_i\cdot t +\tp_i) = \dt{}(m x_i -k\e_{ij}\kx_j)\,,
\ee
or
\bel{3.12} 
\widetilde B_i=p_i\cdot t +\widetilde p_i -mx_i +k\epsilon_{ij}\dot x_j\,.
\ee

After inserting the constraint formula \r{2.10} we derive the following 
conserved generator 
\bel{3.13a}
\widetilde B_i = p_i t - mx_i + 2\widetilde p_i\,,
\ee
Finally, we introduce the boost generators $K_i$ by means of the 
formula\footnotemark[3] 
\bel{3.13b}
\widetilde B_i = p_i t -K_i
\ee
or
\bel{3.14}
 K_i\,=\, mx_i \,-\,2 \tilde p_i\,,
\ee
in consistency with the relation \r{1.6}

Let us recall that the full $D=2$ Galilean algebra is described by the 
generators \r{3.6}, \r{3.8}, \r{3.13a}-\r{3.13b} and the energy operator $H$ 
is given by \r{2.9}. If we use the Dirac brackets \r{2.13}-\r{2.19} it can be 
shown that we obtain the $D=2$ Galilean algebra \r{1.7}, with $E=0$ and 
nonvanishing central charges $M=m\cdot \1$ and $K=2k\cdot \1$.

\subsection*{b) Faddeev-Jackiw method}
Let us write the Lagrangean \r{2.3} as a special case of formula \r{2.23}
\bel{3.15}
L=p_i \dot x_i - k \epsilon_{ij} y_i\dot y_j + \frac{m\vec y^2}{2} -p_i
y_{i}\,.
\ee

The variations corresponding to translations, $O(2)$ rotations and Galilei 
boosts take the form
\bel{3.16}
\ba{rcl}
\delta x_i &=& \delta  \a_i - \e_{ij} x_j \d\a + \d v_i\,t\,,\\[3mm]
\d p_i &=& -\e_{ij}p_i \d \a +m \d v_i\,\\[3mm]
\d y_i &=& -\e_{ij} y_j\d \a +\d v_i\,.
\ea
\ee
Then we obtain 
\bel{3.17}
\d L = \frac d {dt} (m x_i - k\e_{ij}y_j)\d v_i\,.
\ee
Denoting the variations \r{3.16} as $\d{ Y_A} =F_A^r (Y_A,t) \d\, \alpha_r$,
where $Y_A =(x_i,p_i,y_i)$,$\d\a_r= (\d\a_i,\d\a,\d v_i)$ we find
\bel{3.18}
Q^r(t) = P_A(t) F_A^r(Y_A,t)\,,
\ee
where from \r{3.15} it follows that 
\bel{3.19}
P_A = \frac{\p L}{\p\dot Y_A} = (p_i,0, k\e_{ij}y_j)\,.
\ee
Taking into consideration \r{3.17} we see that
\bel{3.20}
\ba{rcl}
P_i&=&p_i\,\\
J &=& \e_{ij}x_i p_j -k y_i^2 \\
K_i &=&p_i t -mx_i + 2k\e_{ij}y_j
\ea
\ee
Together with the formula for the Hamiltonian, which follows from \r{3.15}
\bel{3.21}
H=p_iy_i - \frac{m\vec y^2}2\,,
\ee
and using \r{2.19} one can easily check that
 the formulae \r{3.20}-\r{3.21} lead 
to a realization of the $D=2$ Galilei algebra \r{1.7}
(with E=0)

\sec
\section{Quantization of a free motion with 
    noncommuting space coordinates and internal oscillator modes}

For Galilean systems the position operator $X_i$ can be expressed via 
formula \r{1.3} by the Galilean boost operators $K_i$. Such position 
operators for $D>2$
are commuting. In the $D=2$ case in the presence of two central 
charges ($m\neq0$, $k\neq0$) if we keep the formula \r{1.3} we obtain 
noncommuting position variables. Using \r{3.14} we have
\bel{4.1}
X_i=x_i-\frac2m \tp_i\,.
\ee
If we put $P_i=p_j$ we obtain the standard canonical Poisson bracket
\bel{4.2} 
\{X_i,P_j\}_D=\d_{ij}\,.
\ee
In order to obtain $\{X_i,\tP_j\}_D=0$ we should redefine the second pair of 
momenta as follows
\bel{4.3}
\tP_i =\frac{k}{m} p_i +\e_{ij}\tp_j\,.
\ee
Introducing six phase space variables $Y_A=(X_i,p_i,\tP_i)$ we obtain the 
following symplectic structure
\bel{4.4}
\{Y_A,Y_B\}_D=\Om_{AB}\,,
\ee
where 
\bel{4.5}
\Om=\pmatrix{ \frac{2k}{m\sp2}\e & \1_2 &0 \cr 
-\1_2 &0 &0 \cr
0&0& \frac{k}{2}\e}\,.
\ee

We see from \r{4.4}-\r{4.5} that the parameter $k$ introduces  
noncommutativity in the coordinate sector, in accordance with recent ideas 
of noncommutative geometry (see e.g. [13-15]). One can describe the model 
\r{2.3}
in the Hamiltonian framework using variables $Y_A$. In particular, the 
symmetry generators calculated in Sect. 3 can be expressed as follows
\bl
\beq
\lbl{4.6a} P_i&=&p_i\,,\qquad K_i=m X_i\,,\\
\lbl{4.6b} J&=& -\e_{ij} p_i X_j +\frac{k}{m^2} \vec p{}^2 -\frac1k \vec 
{\tilde P{}}^2\,,\\
\lbl{4.6c} H&=& \frac{ \vec p{}^2}{2m} - \frac{m}{2k^2} \vec{\tilde P{}\sp2}\,.
\eeq
\el
Then the dynamics of the model is described by the following set of 
equations, with $H$ given by (4.6c)
\bel{4.7}
\dot Y_A\,=\,\Omega_{AB}{\partial H\over \partial Y_B},
\ee
or, more explicitly,
\bl
\beq
\lbl{4.8a}\k X_i&=& {2k\over m\sp2}\epsilon_{ij}{\partial H\over \partial X_j}\,+\,{\partial H\over \partial p_i} = \frac{p_i}{m} \,,\\
\lbl{4.8b} \k p_i &=& -{\partial H\over \partial X_i}=0\,,\\
\lbl{4.8c} \ktP_i &=& \frac{k}{2}\epsilon_{ij}{\partial H\over \partial \tilde P_j} =-\frac{m}{2k} \epsilon_{ij}\tilde P_j\,.
\eeq
\el
We see from relations (4.8a,b) that $\ddot X_i=0$, {\it i.e.} that our model describes a free motion
in the noncommutative two-dimensional space, supplemented  by independent
``internal " modes described by 
variables $\tilde P_i$, which can be identified with a standard pair of canonical variables.
Indeed, identifying $\tilde P_1=\sqrt{k\over 2}\tilde x$, $\tilde 
P_2=\sqrt{k\over 2}\tilde p$ 
and introducing oscillator variables
\bel{4.11}
C\,=\,{1\over \sqrt{2}}(\tilde x\,+\,i\,\tilde p),\qquad 
C\sp{\star}\,=\,{1\over \sqrt{2}}(\tilde x\,-\,i\,\tilde p)\ee
we find from \r{4.5} that
\bel{4.12}
\{C,\,C\sp{\star}\}_D\,=\,-i
\ee
and
\bel{4.13}
H\,=\,{\vec p\sp2\over 2m}\,-\,{m\over 4k}(CC\sp{\star}+C\sp{\star}C).
\ee
We see that our model describes a free motion in the noncommutative space 
$(X_1, X_2)$ supplemented by internal degrees of freedom described
by  oscillator modes with negative energies. Indeed, introducing
the correspondence principle
\bel{4.14}
\{A,\,B\}_D\,\rightarrow\,{1\over i\hbar}[\hat A,\,\hat B]
\ee
we obtain from (4.12)
\bel{4.15}
[C,\,C\sp{\dagger}]\,=\,\hbar
\ee
and the spectrum of the Hamiltonian (4.13) is given by
\bel{4.16}
E_{\vec p;n}\,=\,{\vec p\sp2\over 2m}\,-\,{\hbar m\over 4k}(n+\frac12).
\ee

We see that the energy spectrum \r{4.16} is not bounded from below due to 
the presence of the $C$-quanta. The physical space can be defined by 
means of the subsidiary condition
\bel{4.15a}
C\left|ph\right> =0
\ee
Next we shall introduce  interactions which allow us to introduce, 
consistently, the subsidiary condition \r{4.15a}.

\sec
\section{Local potentials in the $D=2$ noncommutative space and the case of 
harmonic forces}

In the previous section we have shown that our model, defined by 
the Lagrangian (2.3), can be decomposed into two decoupled sectors:
\ben
\i[i)]
The external one described by variables $(X_i,\,P_i)$ with  $P_i=p_i$ 
and $X_i$ 
describing noncommuting $D=2$ space coordinates. 
 It appears that our model \r{2.3}
describes a free motion in the 
noncommutative space $X_i$.
\i[ii)] The internal sector which is described by
 auxiliary momenta $\tilde P_i$
which commute with the external variables. The states generated by the 
variables in the internal sector are eliminated by the subsidiary condition 
\r{4.15a}
\een
The model (2.3) describes the one-particle $D=2$ Galilean dynamics and so is fixed uniquely by the Galilean invariance with $k\ne0$ and $E=0$
(see (1.7))\footnotemark[4]. In order to add to the free Lagrangian (2.3) a
potential energy term, consistent with Galilean invariance, we have 
to consider the two-body particle dynamics. Denoting by $x_{i;1}$, $x_{i;2}$ ($i=1,2)$ 
the positions of two point particles we consider the following
Lagrangian\footnotemark[5]
\bel{5.1}
L_{1+2}=L_{0;1}+L_{0;2}\,-\,V(x_{i;1}-x_{i;2},\,\dot x_{i;1}-\dot x_{i;1}),
\ee
where $V$ is a scalar with respect to the $D=2$ space rotations $O(2)$ 
and  is, by construction, invariant under translations and Galilean 
boosts. $L_{0;r}$ are given by (for simplicity we put $\tilde m=m_1=m_2$
and $\tilde k=k_1=k_2$)
\bel{5.2}
L_{0;r}\,=\,{\tilde m\over 2}\dot x_{i;r}\sp2\,-\,\tilde k\e_{ij}\dot x_{i;r}\ddot x _{j;r}.
\ee
If we define the centre-of-mass (CM) and relative coordinates
\bel{5.3}
x_i:=\,x_{i;1}-x_{i;2},\qquad R_i:=\frac12 (x_{i;1}+x_{i;2})
\ee
we can rewrite (5.1) as
\bel{5.4}
L_{1+2}=L_{CM}+L'
\ee
where $(M=2\tilde m$, $K=2\tilde k$; $m=\frac{\tilde m}{2}$, $k=\frac{\tilde k}{2})$

\bl
\beq
\lbl{5.5a}L_{CM}\,=\,{M\over 2}\dot R_i\sp2-K\e_{ij}\dot R_i\ddot R_j \,,\\
\lbl{5.5b} L'={m\over 2}\dot x_i\sp2\,-\,k\,\e_{ij}\dot x_i\ddot x_j\,-\,V(x_i,\dot x_i)\,.
\eeq
\el

The global CM motion described by (5.5a) has exactly the structure of the one-particle
dynamics discussed in sections 2-4. In the following we shall study
the dynamics of the relative two-particle motion described by the Lagrangian 
(5.5b). We postulate that
the Hamiltonian obtained from the Lagragian (5.5b) should also split into 
the sum
\bel{5.7}
H\,=\,H\sp{(ext)}(P,X)\,+\,H\sp{(int)}(\tilde P)\,,
\ee
where $H\sp{(int)}(\tilde P)$ is the free internal oscillator Hamiltonian
(see \r{4.11}).
We shall consider here the interactions $V(x_i,\dot x_i)$ which do
 not modify the  choice
of the internal Hamiltonian and add to the free external Hamiltonian an arbitrary potential $U(X)$
\bel{5.8}
H\sp{(ext)}(P,X)\,=\,{P\sp2\over 2m}\,+\,U(X).
\ee
For this purpose we will take
\bel{5.9}
V(x_i,\dot x_i)\,=\,U(x_i-{2k\over m}\epsilon_{ij}\dot x_j).
\ee
Using 
\bel{5.9a}
p_i=m \dot x_i - 2k\e_{ij}\ddot x_j - \frac{2k}m \e_{ij} \frac{\p U}{\p 
x_j}\,,\qquad \tilde p_i=k\e_{ij}\dot x_j\,,
\ee
and \r{4.1}, \r{4.3}, \r{5.9a}
 we get
\bel{5.10}
H=\dot x_i p_i+\ddot x_i\tilde p_i-\L={\vec p\sp2\over 2m}-{m\over 2k\sp2}\vec{\tilde P}\sp2+U(X_i).
\ee
We see therefore that our particular velocity-dependent interaction (5.8) 
leads to local interactions involving noncommutative variables.

The property that quantum mechanical models on noncommutative spaces can be
transformed into standard but complicated models on commuting spaces
is known from the studies of quantum deformed models of
quantum mechanics (see e.g. [17,18]).
Our model provides one more example of such a construction. In the simplest case
one can assume that the potential $U$ is quadratic and so consider the following
form of the noncommutative oscillator Lagrangian
\bel{5.11}
L_{osc}={m\dot x_i\sp2\over 2}-k\epsilon_{ij}\dot x_i\ddot x_j-{m\om\sp2\over 2}
(x_i-{2k\over m}\epsilon_{ij}\dot x_j)\sp2
\ee
which contains the Chern-Simons term.
The equations (2.6) are now generalised to
\bel{5.12}
(\d_{ij}-{2k\over m}\e_{ij}{d\over dt})(m\ddot x_j-2k\om\sp2\e_{jl}\dot x_l+m\om\sp2 x_j)=0
\ee

Introducing noncommutative coordinates (4.1) and the modified auxiliary
momenta (4.3) we obtain (putting $P_i=p_i$) the expected special form of 
\r{5.8}
\bel{5.15}
H_{osc}\sp{(ext)}(P,X)\,=\,{\vec P\sp2\over 2m}\,+\,{m\om\sp2\over 2}\vec X\sp2.
\ee
Hamilton's equations \r{4.7} then give us
\bel{5.16}
\ba{c}
\dsp \dot X_i\,=\,2{k\om\sp2\over m}\e_{ij}X_j\,+\,{1\over m}P_i\\[3mm]
\dsp \dot P_i\,=\,-m\om\sp2 X_i.
\ea
\ee
Thus we obtain the following equation for our noncommutative $X_i$:
\bel{5.17}
\ddot X_i\,-\,2 {k\om\sp2\over m}\e_{ij}\dot X_j\,+\,m\om\sp2 X_i\,=\,0.
\ee
We note that the velocity dependent term in \r{5.17} is due to the 
noncommutativity of space coordinates $X_i$ (see (1.9)).
If we now introduce, in the standard way, the oscillator variables
\bl
\beq
\label{5.18a}A_i\,:=\,\sqrt{m\om\over 2}X_i\,+\,i\sqrt{1\over 2m\om}P_i\\
\label{5.18b}A_i\sp{*}\,:=\,\sqrt{m\om\over 2}X_i\,-\,i\sqrt{1\over 2m\om}P_i
\eeq
\el
we obtain
\bel{5.19}
H_{osc}\sp{(ext)}\,=\,{\om\over 2}(A_iA_i\sp{*}+A_i\sp{*}A_i).
\ee
Calculating the Dirac brackets (see \r{4.5}) for the oscillator variables 
\r{5.18a} - \r{5.18b} and quantizing by 
the substitution $\{\,.\,,\,\,.\,\}_D\rightarrow 
{1\over i\hbar}[\,.\,,\,\,.\,]$ we 
obtain\bel{5.20}
\ba{rcl}
[A_i,\,A_j\sp{\dagger}]&=&\dsp
\hbar \d_{ij}\,+\,{i\hbar \om k\over m}\e_{ij}\\[3mm]
[A_i,\,A_j]&=&\dsp [A_i\sp{\dagger},\,A_j\sp{\dagger}]\,=\,{i\hbar 
\om k\over m}\e_{ij}.\ea\ee
The $k$-deformation of the harmonic oscillator is obtained therefore by 
the deformation of the Heisenberg commutation relations, describing 
modified equal time oscillator algebra \r{5.20}. 
In principle one can quantize the 
$k$-deformed oscillator by introducing a $k$-deformed Fock space of the 
modified oscillator variables \r{5.20}. Here we shall however 
solve the model by introducing the following commuting space coordinates
\bel{5.19n}
\hat X_i = X_i +\frac k{m^2} \e_{ij} P_j = x_i - \frac{2}m \tilde p_i + 
\frac k{m^2}\e_{ij}p_j\,,
\ee
which satisfy two conditions:
\ben
\i[i)] the standard Poisson brackets are valid
\bel{5.20n}
\{ \hat X_i, P_j \} = \d_{ij}\,,\qquad \{ \hat X_i, \hat X_j \} = 0\,.
\ee
\i[ii)] the Poissons bracket with internal symmetry variables  vanish, 
i.e.\ $\{\hat X_i, \tilde P_j\} =0$.
\een

 In this way, following [10], we find that \r{5.15} gives us
\bel{5.21}
\ba{rcl}
H_{osc}\sp{(ext)}&=&\dsp {\vec p\sp2\over 2m}\,+\,{m\om\sp2\over 2}(\hat 
X_i-{k\over m\sp2}\e_{ij}P_j)\sp2\,=\\[3mm]
&=&\dsp {p\sp2\over 2\tilde 
m}\,+\,{\tilde m\tilde\om\sp2\over 2}\hat X_i\sp2\,-\,{k\om\sp2\over 
m}\,J\sp{(ext)}\\
\ea
\ee
where
\bl
\beq
\label{5.22a}\tilde m=m(1+\om\sp2{k\sp2\over m\sp2})\sp{-1}\\
\label{5.22b}\tilde \om\sp2\,=\,\om\sp2(1+{k\sp2\om\sp2\over m\sp2})
\eeq
\el
and where, as is clear from (4.6b),
\bel{5.23}
J\sp{(ext)}\,=\,\e_{ij}X_iP_j\,+\,{k\over m\sp2}\vec p\sp2\,=\,\e_{ij}\hat X_iP_j
\ee
describes the $O(2)$ angular momentum for the external dynamics.

The first part of the Hamiltonian \r{5.21} is the standard oscillator. 
If we introduce the standard quantized oscillator variables 

\bel{5.25}
\ba{c}
\dsp a_i\,:=\,\sqrt{\tilde m\tilde\om\over 2\hbar}\hat X_i\,+\,i\sqrt{1\over 
2\tilde m\tilde\om\hbar}P_i\\[3mm]
\dsp a_i\sp{\dagger}\,:=\,\sqrt{\tilde 
m\tilde\om\over 2\hbar}\hat X_i\,-\,i\sqrt{1\over 2\tilde 
m\tilde\om\hbar}P_i\\
\ea\ee
satisfying 
\bel{5.26}
[a_i,\,a_j\sp{\dagger}]\,=\,\d_{ij}
\ee
 we find that
\bel{5.27}
H_{osc}\sp{(ext)}\,=\,H_{osc}\sp{(0)}\,-\,\gamma\,J\sp{ext},\qquad \gamma={k\om\sp2\over m},
\ee
where
\bl
\beq
\label{5.28a} H_{osc}\sp{(0)}\,=\,\hbar \tilde \om(a_i\sp{\dagger}a_i+1)
=\hbar\tilde \om(N+\frac12)\\
\label{5.28b} J\sp{ext}\,=\,i\hbar (a_2\sp{\dagger}a_1-a_1\sp{\dagger}a_2).
\eeq
\el
As is well known, using the operators (5.25) and the $2\times 2$ 
Pauli matrices $\sigma_r\,(r=1,2,3)$, we can construct the following
 $SU(2)$ Lie algebra
generators\footnotemark[6]
\bel{5.29}
J_r=\frac12 a_i\sp{\dagger}(\sigma_r)_{ij}a_j.
\ee
The $SU(2)$ Casimir $J\sp2=J_rJ_r$ is related to the number operator (see (5.27a)) by
\bel{5.30}
J\sp2\,=\,{N\over 2}({N\over 2}+1).
\ee
Furthermore
\bel{5.31}
J\sp{ext}=2\hbar J_2.
\ee

Let us consider now the common eigenstates of $N$ and $J_2$ (see also [9])
\bl
\beq
\label{5.32a} N|n;l>=n|n;l>\qquad n\in N\\
\label{5.32b} J_2|n;l>={l\over 2}|n;l>\qquad l\in Z.
\eeq
\el 
We see from (5.29), however, that in the oscillator representation the number $n$ plays the role of the half of the angular momentum 
eigenvalue. Following the standard procedure in quantum mechanics we see that
we have the restriction
\bel{5.33}
|l|\le n.
\ee

From (5.26), (5.27a,b), (5.30) and (5.31a,b) we obtain
\bel{5.34}
H_{osc}\sp{(ext)}|n;l>=E_{n,l}|n;l>,
\ee
where
\bel{5.35}
E_{n,l}=\hbar \tilde \om(n+1)-\hbar {k\om\sp2\over m}l.
\ee
Using (5.32) we obtain the following lower bound on the energy spectrum
\bel{5.36}
E_{n,l}\ge \hbar \tilde \om +\hbar n(\tilde \om -{k\om\sp2\over m}).
\ee

From (5.22b) we get
\bel{5.37}
\tilde \om -{k\om\sp2\over m}\,=\,\om\bigl(\sqrt{1+{k\sp2\om\sp2\over m\sp2}}
-{k\om \over m}\bigr)>0.
\ee
We see that the energy spectrum (5.35) is positive.

In order to describe states (5.31a-b) one should consider the oscillators $a
_i$ as  $SU(2)$ spinors and rotate them around the first axis by an angle
${\pi\over 4}$. Introducing the following unitary transformation 
\bel{5.38}
\tilde a_i=U_{ij}a_j,
\ee
where 
\bel{5.39}
U=\bigl[ exp{i\pi\over 4}(\sigma_1)\bigr]_{ij}\,=\,{1\over \sqrt{2}}(1+i\sigma_1)
\ee
we find that
\bel{5.40}
J_2\,=\,{1\over 2}(\tilde a_1\sp{\dagger}\tilde a_1
-\tilde a_2\sp{\dagger}\tilde a_2).
\ee
If we now introduce the states
\bel{5.41}
|n_1,n_2>\,=\,{1\over \sqrt{n_1!}}{1\over \sqrt{n_2!}}(\tilde a_1\sp{\dagger})\sp{n_1}(\tilde a_2\sp{\dagger})\sp{n_2}|0>
\ee
where
\bel{5.42}
\tilde a_i|0>=0
\ee
we obtain the following formula for the eigenstates (5.33)
\bel{5.43}
|n;l>\,=\,|\frac12 (n+l),\frac12(n-l)>.
\ee

We would like to make the following two additional remarks
\ben
\i[i)]
The Lagrangian (5.5a) can also be discussed for nonharmonic potentials $U(X_i)$.
For example, one can assume that $U(X_i)={\lambda\over 4}(\vec{X\sp2})\sp2$.
It is easy to see that such a model with noncommutative quartic interaction
will lead, in the commuting space, to a Hamiltonian with the 
generalised kinetic term 
\bel{5.44}
{\vec p\sp2\over 2m}\,\rightarrow\,
{\vec p\sp2\over 2m}+{\tilde \lambda\over 4}(\vec p\sp2)\sp2,\qquad \tilde \lambda={\lambda k\sp4\over m\sp8}.
\ee
Such a model is currently under consideration.
\i[ii)]
One can also ask if it is possible to generalize the free oscillator Hamiltonian
$H\sp{(int)}$ (see \r{5.7}).
It can be shown that such a generalization is possible only if we can introduce
 terms with second order time derivatives
which are different from the Chern-Simons-like term in \r{2.3}. It appears, 
however,that in such a case, the constraints \r{2.10} are not valid and the 
separationinto ``external" and ``internal" degrees of freedom looses its 
meaning.Moreover, for the interacting ``internal" degrees of freedom the 
negativemetric states cannot be consistently eliminated by the imposition of 
a subsidiary condition, and the Faddeev-Jackiw method cannot be applied.
\een

\sec
\section{Concluding Remarks}
\phantom{aa}

\hskip 0.5cm  In sect. 2-4 we have presented a one-particle 
model with higher order derivatives, which provides a 
dynamical interpretation of the second central extension of $D=2$ Galilei 
algebra. The model can be interpreted as describing a free motion 
in the $D=2$ space with noncommuting coordinates and internal structure 
described by oscillator modes with negative energies. Further, in sect. 5,
we have also considered  a two-particle
Galilean-invariant dynamics with relative motion
described by a model with a potential depending on
noncommuting coordinates. It appears that such models are obtained 
if our primary Lagrangian contains suitably chosen velocity-dependent
interactions. In particular, we have fully discussed the case of a harmonic 
oscillator in noncommutative space. The modification
 due to the noncommutativity introduces  a bilinear $SU(2)$-breaking term
into
the Hamiltonian of
 a two-dimensional oscillator.

It appears that the dynamics in the models considered in Sect.\ 5 can 
be separated into two independent sectors - describing external and internal
dynamics. The external dynamics describes the quantum mechanics in the $D=2$ 
space with noncommuting coordinates. The internal dynamics, in the 
presence of particularly chosen interaction in the external sector, is 
described by free oscillators (4.15). 

From the general framework for 
higher order Lagrangians (see eg [21]) it can be deduced
 that in our model there exist states 
which, after quantization, are endowed with an indefinite metric. These states 
are generated by the internal oscillator variables  and are 
eliminated from the physical spectrum by the imposition of the subsidiary condition
\r{4.15a}. 
In the  case of the interaction described by the potential (5.9)
we obtain from the Lagrangian (5.5b) the following equations of motion
\bel{6.3}
\hat O_{ij}(m\ddot x_j+{\partial U\over \partial x_j})=0,
\ee
where
\bel{6.4}
\hat O_{ij}=(\d_{ij}-{2k\over m}\e_{ij}{d\over dt}).
\ee
The subsidiary condition \r{4.15a} eliminating the states with negative 
metric can be expressed in the following way
\bel{6.5}
 [\ddot x_i(t)+{1\over m}{\partial U\over \partial x_i}]|Ph>=0
\ee
{\it i.e.} in the physical sector we retain only the reduced dynamics in 
the external sector. The factorization of the Euler-Lagrange equations (6.3)
shows that the operator $\hat O_{ij}$ describing the modes
which carry indefinite metric does not depend on the interaction
term and the relation \r{6.5} is equivalent to the subsidiary condition
\r{4.15a}.

 In summary, one can treat the presented model as an explicit 
realization of a theory with higher derivatives, with interesting symmetry 
properties, 
constraints and several symplectic structures. 
The model also illustrates  very well the Faddeev-Jackiw technique of 
quantization.
In our view it is also important that the model provides
an example of a Lagrangian dynamics which can be expressed 
equivalently in terms of commuting and noncommuting position
variables.
Moreover, although our model contains higher order derivatives,
even in the interacting case, the ghost problem of higher order Lagrangian theories can be solved. The unphysical features
(negative energies, negative metric states) which are linked to
 the introduction
of a noncommutative structure are described by free modes and can be made harmless by the imposition
of a suitable subsidiary condition eliminating negative metric states.

Finally we would like to add that since the appearance of recent models of strings  with substructure described by the so-called 
0-branes (see e.g. [28,29])
the (2+1) dimensional Galilei-invariant systems
have become more important. Possible links with such applications 
should  also be considered.
\vskip 0.5cm
{\bf Acknowledgments}. \hfil\break
\vskip 0.5cm
One of the authors (JL) would like to thank the Department of Mathematical Sciences of the University of Durham for its warm hospitality and the EPSRC for financial support. 
He also acknowledges partial support from the KBN grant 2P30208706. 

The authors would like to thank Roman Jackiw for pointing out that our 
model can be quantized in an easy way by the geometric method presented in 
[13, 14]. The authors would also like to acknowledge  Jose A. de Azcarraga 
and E.H. de Groot for helpful comments.

\sec
\renewcommand{\theequation}{A.\arabic{equation}}
\annexe{1}{Appendix}


\par

In this appendix we prove the following\hfil\break
{\bf Theorem}. The most general one-particle Lagrangian, which is at most linearly dependent on $\ddot x$, leading to the Euler-Lagrange equations of motion 
which are covariant with respect to the $D=2$ Galilei group, is given, up to
gauge transformations, by
\be
L(x,\dot x, \ddot x,t)\,=\,\frac m2 \dot x_i\dot x_i\,-\,k\,\epsilon_{ij}
\dot x_i\ddot x_j
\ee
with $m$ and $k$ constant.

\noindent {\bf Proof}
\begin{itemize}
\item{(i)} Covariance of the equations of motion, with respect to the $D=2$ Galilei group 
is equivalent to the statement that the Euler variation $f_i$ is independent
of $t$, $x$ and $\dot x$ and transforms, under space rotations, as an
$i\sp{th}$ component of a vector (see [23]), {\it i.e.}
\bel{A.2}
\pa{L}{x_i}-\dt{}\pa{L}{\kx_i}+\frac{d^2}{dt^2}\pa{L}{\kkx_i}=f_i(\ddot x,
\ddd,\dddd).
\ee
\item{(ii)} If we now suppose that $L$ is at most linearly dependent on $\ddot x$
\bel{A.3}
L(x,\dot x,\ddot x,t)\,=\,L_1(x,\dot x,t)\,+\,L_{2i}(x,\dot x,t)\ddot x_{i}
\ee
we conclude from (A.2) that $f_i$ does not depend on $\dddd$ and is linearly dependent on $\ddd$
\bel{A.4}
f_i(\ddot x,\ddd)\,=\,f_{1i}(\ddot x)\,+\,f_{2ij}(\ddot x) {\ddd_j}
\ee
with
\bel{A.5}
f_{2ij}(\ddot x)\,=\,{\partial L_{2i}\over \partial \dot x_j}
-{\partial L_{2j}\over \partial \dot x_i}.
\ee
Because the r.h.s. of (A.5) is an antisymmetric tensor independent 
of $\ddot x$, we have
\bel{A.6}
f_{2ij}(\ddot x)\,=\,2k\epsilon_{ij}
\ee
with $K$=constant (the factor 2 is a matter of convenience).

Putting (A.6) back ito (A.5) we conclude that
\bel{A.7}
L_{2i}(x,\dot x,t)=L_{20i}(x,t)\,+\,L_{21ij}(x,t)\dot x_j\,+\,k\epsilon_{ij}\dot x_j
\ee
with $L_{21ij}$ being a symmetric tensor.\hfil\break
Therefore we have as an intermediate result
\bel{A.8}
\ba{c}
L(x,\dot x,\ddot x,t)\,=\,L_1(x,\dot x,t)\,+\,L_{20i}(x,t)\ddot x_i \\
+k\epsilon_{ij}\,\ddot x_i\,\dot x_j\,+\,L_{21ij}(x,t)\,\ddot x_i\,\dot x_j.
\ea
\ee
\item{(iii)} We now perform a gauge transformation
\bel{A.9}
L\,=\,\tilde L\,+\,{d \over dt}\phi(x,\dot x,t)
\ee
with
$\phi:=L_{20i}(x,y)\dot x_i\,+\,\frac12 L_{21ij}(x,t)\dot x_i\dot x_j.$

Then $\tilde L$ reads
\bel{A.10}
\tilde L(x,\dot x,\ddot x,t)\,=\,L_0(x,\dot x,t)\,+\,k\,\epsilon_{ij}\,\ddot x_i\dot x_j
\ee
with
\bel{A.11}
\ba{c}
L_0\,:=\,L_1(x,\dot x,t)\,-\,({d\over dt}L_{20i}(x,t))\dot x_i\\
-\frac 12({d\over dt}L_{21ij}(x,t))\dot x_i\dot x_j.
\ea
\ee
Because the Euler-variation is invariant with respect to the gauge
transformation (A.9) we find from (A.2) and (A.4) that
\bel{A.12}
{\partial L_0(x,\dot x,t)\over \partial x_i} -{d\over dt} {\partial L_0(x,\dot x,t)\over \partial \dot x_i}\,=\,f_{1i}(\ddot x).
\ee
But (A.12) is a well known text-book problem (cp. Landau - Lifshitz, Vol.I)
with the solution
\bel{A.13}
L_o(x,\dot x,t)\,=\,\frac m2 \dot x_i\dot x_i\,+\,\frac d{dt}\psi(x,t).
\ee
\item{(iv)}
Combining (A.8), (A.9) and A(13) we arrive at (A.1) up to a gauge 
transformation.
\end{itemize}
\sec
\renewcommand{\theequation}{B.\arabic{equation}}

\section*{Footnotes}
\ben
\i[1.] We consider here $\hbar=c=1$, i.e\ the mass dimensions of space and 
time coordinates are the same and equal to $-1$.
\i[2.] The noncommutativity (1.9) (though it describes
the $D=2$ case) resembles the four-dimensional noncommutative
structure proposed in [17], where the space-time coordinates also commute
to a number.
\i[3.]For convenience we have changed the overall sign of the modified bost
generators. 
\i[4.] Adding potential $V(x)$ to a one-particle dynamics leads to the broken 
translational invariance and we obtain $[P_i,\,H]={\partial \over \partial 
x_i}V(x)\ne 0$. In the $D=3$ case the modification of the Galilei 
algebra obtained by requiring that the one-particle dynamics is described 
by a harmonic oscillator was first discussed by Sudbery [26].
\i[5.] Following 
[25] one can show that the $D=2$ Lagrangian $L_{1+2}(x_{i;r},\dot 
x_{i;r},\ddot x_{i;r})$ $(i=1,2)$; $(r=1,2)$ for the interacting identical 
point particles is the most general one which\ben
\i[i)] contains only linear acceleration-dependent terms
\i[ii)] the potential $V$ depends only on coordinates $x_{i;r}$ and velocities
$\dot x_{i;r}$
\i[iii)] it leads to the Euler-Lagrange equations of motion which are form-invariant with respect to the $D=2$ Galilei transformations (1.1)
\een
\i[6.] This is the so-called Schwinger-Jordan representation (see [27]).
\een
\newpage
\def\b{\bibitem}

\end{document}